\documentclass[preprintnumbers,amsmath,amssymbm,prl]{revtex4}
\usepackage{epsfig}
\usepackage{graphicx}

\begin{document}
\title{Quasinormal resonances of near-extremal Kerr-Newman black holes}
\author{Shahar Hod}
\address{The Ruppin Academic Center, Emeq Hefer 40250, Israel}
\address{ }
\address{The Hadassah Institute, Jerusalem 91010, Israel}
\date{\today}

\begin{abstract}
\ \ \ We study {\it analytically} the fundamental resonances of
near-extremal, slowly rotating Kerr-Newman black holes. We find a
simple analytic expression for these black-hole quasinormal
frequencies in terms of the black-hole physical parameters:
$\omega=m\Omega-2i\pi T_{BH}(l+1+n)$, where $T_{BH}$ and $\Omega$
are the temperature and angular velocity of the black hole. The mode
parameters $l$ and $m$ are the spheroidal harmonic index and the
azimuthal harmonic index of a co-rotating mode, respectively. This
analytical formula is valid in the regime $\Im\omega\ll \Re\omega\ll
M^{-1}$, where $M$ is the black-hole mass.
\end{abstract}
\bigskip
\maketitle


The statement that black holes have no hair was introduced by
Wheeler \cite{Wheeler} in the early 1970's. The various no-hair
theorems state that the external field of a dynamically formed black
hole (or a perturbed black hole) relaxes to a Kerr-Newman spacetime,
characterized solely by three parameters: the black-hole mass,
charge, and angular momentum. This implies that perturbation fields
left outside the black hole would either be radiated away to
infinity, or be swallowed by the black hole.

This relaxation phase in the dynamics of perturbed black holes is
characterized by `quasinormal ringing', damped oscillations with a
discrete spectrum (see e.g. \cite{Nollert1} for a detailed review).
At late times, all perturbations are radiated away in a manner
reminiscent of the last pure dying tones of a ringing bell
\cite{Press,Cruz,Vish,Davis}. Being the characteristic `sound' of
the black hole itself, these free oscillations are of great
importance from the astrophysical point of view. They allow a direct
way of identifying the spacetime parameters (the mass, charge, and
angular momentum of the black hole). This fact has motivated a
flurry of research during the last four decades aiming to compute
the quasinormal mode (QNM) spectrum of various types of black-hole
spacetimes \cite{Nollert1}.

The dynamics of black-hole perturbations is governed by the
Regge-Wheeler equation \cite{RegWheel} in the case of a spherically
symmetric Schwarzschild black hole, and by the Teukolsky equation
\cite{Teu} for rotating Kerr-Newman spacetimes. The black hole QNMs
correspond to solutions of the wave equations with the physical
boundary conditions of purely outgoing waves at spatial infinity and
purely ingoing waves crossing the event horizon \cite{Detwe}. Such
boundary conditions single out a discrete set of black-hole
resonances $\{\omega_n\}$ (assuming a time dependence of the form
$e^{-i\omega t}$). (In analogy with standard scattering theory, the
QNMs can be regarded as the scattering resonances of the black-hole
spacetime. They thus correspond to poles of the transmission and
reflection amplitudes of a standard scattering problem in a
black-hole spacetime.)

Since the perturbation field can fall into the black hole or radiate
to infinity, the perturbation decays and the corresponding QNM
frequencies are {\it complex}. It turns out that there exist an
infinite number of quasinormal modes, characterizing oscillations
with decreasing relaxation times (increasing imaginary part)
\cite{Leaver}. The mode with the smallest imaginary part (known as
the fundamental mode) determines the characteristic dynamical
timescale $\tau$ for generic perturbations to decay.

In this work we determine analytically the {\it fundamental}
(least-damped) resonant frequencies of rotating Kerr-Newman black
holes. (For a recent progress in the study of the {\it
highly}-damped resonances, see \cite{KeshHod,Ber3}.) The spectrum of
quasinormal resonances can be studied analytically in the slow
rotation, near-extremal limit $(M^2-Q^2-a^2)^{1/2}\ll a\ll M$, where
$M$, $Q$, and $a$ are the mass, charge, and angular momentum per
unit mass of the black hole, respectively. In order to determine the
black-hole resonances we shall analyze the scattering of massless
scalar and neutrino waves in the Kerr-Newman spacetime
\cite{Noteno,Chan}. The dynamics of a perturbation field $\Psi$ in
the Kerr-Newman spacetime is governed by the Teukolsky equation
\cite{Teu}. One may decompose the field as (we use natural units in
which $G=c=\hbar=1$)

\begin{equation}\label{Eq1}
\Psi_{slm}(t,r,\theta,\phi)=e^{im\phi}S_{slm}(\theta;a\omega)\psi_{slm}(r;a\omega)e^{-i\omega
t}\ ,
\end{equation}
where $(t,r,\theta,\phi)$ are the Boyer-Lindquist coordinates,
$\omega$ is the (conserved) frequency of the mode, $l$ is the
spheroidal harmonic index, and $m$ is the azimuthal harmonic index
with $-l\leq m\leq l$. The parameter $s$ is called the spin weight
of the field, and is given by $s=\pm {1\over 2}$ for massless
neutrino perturbations, and $s=0$ for scalar perturbations. (We
shall henceforth omit the indices $s,l,m$ for brevity.) With the
decomposition (\ref{Eq1}), $\psi$ and $S$ obey radial and angular
equations, both of confluent Heun type \cite{Heun,Flam}, coupled by
a separation constant $A(a\omega)$.

The angular functions $S(\theta;a\omega)$ are the spin-weighted
spheroidal harmonics which are solutions of the angular equation
\cite{Teu,Flam}

\begin{equation}\label{Eq2}
{1\over {\sin\theta}}{\partial \over
{\partial\theta}}\Big(\sin\theta {{\partial
S}\over{\partial\theta}}\Big)+\Big[a^2\omega^2\cos^2\theta-2a\omega
s\cos\theta-{{(m+s\cos\theta)^2}\over{\sin^2\theta}}+s+A\Big]S=0\ .
\end{equation}
The angular functions are required to be regular at the poles
$\theta=0$ and $\theta=\pi$. These boundary conditions pick out a
discrete set of eigenvalues labeled by an integer $l$. In the
$a\omega\ll 1$ limit these angular functions become the familiar
spin-weighted spherical harmonics with the corresponding angular
eigenvalues $A=l(l+1)-s(s+1)+O(a\omega)$ \cite{Notealm}.

The radial Teukolsky equation is given by

\begin{equation}\label{Eq3}
\Delta^{-s}{{d}
\over{dr}}\Big(\Delta^{s+1}{{d\psi}\over{dr}}\Big)+\Big[{{K^2-2is(r-M)K}\over{\Delta}}
-a^2\omega^2+2ma\omega-A+4is\omega
r\Big]\psi=0\ ,
\end{equation}
where $\Delta\equiv r^2-2Mr+Q^2+a^2$ and $K\equiv
(r^2+a^2)\omega-am$. The zeroes of $\Delta$, $r_{\pm}=M\pm
(M^2-Q^2-a^2)^{1/2}$, are the black hole (event and inner) horizons.

For the scattering problem one should impose physical boundary
conditions of purely ingoing waves at the black-hole horizon and a
mixture of both ingoing and outgoing waves at infinity (these
correspond to incident and scattered waves, respectively). That is,

\begin{equation}\label{Eq4}
\psi \sim
\begin{cases}
e^{-i\omega y}+{\mathcal{R}}(\omega)e^{i \omega y} & \text{ as }
r\rightarrow\infty\ \ (y\rightarrow \infty)\ ; \\
{\mathcal{T}}(\omega)e^{-i (\omega-m\Omega)y} & \text{ as }
r\rightarrow r_+\ \ (y\rightarrow -\infty)\ ,
\end{cases}
\end{equation}
where the ``tortoise" radial coordinate $y$ is defined by
$dy=[(r^2+a^2)/\Delta]dr$. Here $\Omega\equiv {a\over {r^2_++a^2}}$
is the angular velocity of the black-hole horizon. The coefficients
${\cal T}(\omega)$ and ${\cal R}(\omega)$ are the transmission and
reflection amplitudes for a wave incident from infinity. The
discrete black-hole resonances are the poles of these transmission
and reflection amplitudes. (The pole structure reflects the fact
that the QNMs correspond to purely outgoing waves at spatial
infinity.) These resonances determine the ringdown response of a
black hole to outside perturbations.

The transmission and reflection amplitudes satisfy the usual
probability conservation equation $|{\cal T}(\omega)|^2+|{\cal
R}(\omega)|^2=1$. The calculation of these scattering amplitudes in
the low frequency limit, $\Im\omega\ll \Re\omega\ll M^{-1}$, is a
common practice in the physics of black holes, see e.g.
\cite{Chan,Page,Hodcen} and references therein. Define

\begin{equation}\label{Eq5}
x\equiv {{r-r_+}\over {r_+-r_-}}\ \ ;\ \
\varpi\equiv{{\omega-m\Omega}\over{4\pi T_{BH}}}\ \ ;\ \ k\equiv
\omega(r_+-r_-)\  ,
\end{equation}
where $T_{BH}={{(r_+-r_-)}\over{4\pi(r^2_++a^2)}}$ is the
Bekenstein-Hawking temperature of the black hole. Then a solution of
Eq. (\ref{Eq3}) obeying the ingoing boundary conditions at the
horizon ($r\to r_+$, $kx\ll 1$) is given by \cite{Morse,Abram}

\begin{eqnarray}\label{Eq6}
\psi=x^{-s-i\varpi}(x+1)^{-s+i\varpi}
{_2F_1}(-l-s,l-s+1;1-s-2i\varpi;-x) \ ,
\end{eqnarray}
where $_2F_1(a,b;c;z)$ is the hypergeometric function. In the
asymptotic ($r\gg M$, $x\gg |\varpi| +1$) limit one finds the
solution \cite{Morse,Abram}

\begin{eqnarray}\label{Eq7}
\psi&=&C_1e^{-ikx}x^{l-s}{_1F_1}(l-s+1;2l+2;2ikx)\nonumber
\\&& +C_2e^{-ikx}x^{-l-s-1}{_1F_1}(-l-s;-2l;2ikx)\ ,
\end{eqnarray}
where $_1F_1(a;c;z)$ is the confluent hypergeometric function. The
coefficients $C_1$ and $C_2$ can be determined by matching the two
solutions in the overlap region $|\varpi|+1\ll x\ll 1/k$. This
yields

\begin{equation}\label{Eq8}
C_1={{\Gamma(2l+1)\Gamma(1-s-2i\varpi)}\over
{\Gamma(l-s+1)\Gamma(l+1-2i\varpi)}}\
 ,
\end{equation}
and
\begin{equation}\label{Eq9}
C_2={{\Gamma(-2l-1)\Gamma(1-s-2i\varpi)}\over
{\Gamma(-l-s)\Gamma(-l-2i\varpi)}}\ .
\end{equation}
Finally, the asymptotic form of the confluent hypergeometric
functions \cite{Morse,Abram} can be used to write the solution in
the form given by Eq. ({\ref{Eq4}). After some algebra one finds

\begin{equation}\label{Eq10}
|{\cal T}(\omega)|^2=\Re
\Big\{\Big[{{(l-s)!(l+s)!}\over{(2l)!(2l+1)!}}\Big]^2{{\Gamma(l+1-2i\varpi)}\over{\Gamma(-l-2i\varpi)}}(2ik)^{2l+1}\Big\}\
,
\end{equation}
for the transmission probability.

The quasinormal frequencies are the scattering resonances of the
black-hole spacetime. They thus correspond to poles of the
transmission and reflection amplitudes. Taking cognizance of Eq.
(\ref{Eq10}) and using the well-known pole structure of the Gamma
functions \cite{Abram}, one finds the resonance condition
$l+1-2i\varpi=-n$, where $n\geq 0$ is a non-negative integer. This
yields a simple formula for the black-hole resonances:

\begin{equation}\label{Eq11}
\omega=m\Omega-2i\pi T_{BH}(l+1+n)\  ,
\end{equation}
in the near-extremal limit. It is worth emphasizing again that this
formula is valid in the $\Im\omega\ll\Re\omega\ll M^{-1}$ regime.
This requires $(M^2-Q^2-a^2)^{1/2}\ll a\ll M$ and $m>0$
\cite{Notevalid}.

In summary, we have studied analytically the quasinormal mode
spectrum of near-extremal, slowly rotating Kerr-Newman black holes.
It was shown that the fundamental resonances can be expressed in
terms of the black-hole physical parameters: the temperature
$T_{BH}$, and the horizon angular velocity $\Omega$.

The fundamental resonances are expected to dominate the relaxation
dynamics of a perturbed black-hole spacetime. Taking cognizance of
Eq. (\ref{Eq11}), one realizes that in the near-extremal limit
$\Im\omega$ approaches zero linearly with the black-hole temperature
$T_{BH}$ for all modes co-rotating with the black hole (i.e., modes
having $m>0$). We therefore conclude that the characteristic
relaxation timescale $\tau \sim 1/\Im\omega$ of the black hole is of
the order of $O(T_{BH}^{-1})$ \cite{Notedim}. It is worth
emphasizing that this result, $\tau \sim T_{BH}^{-1}$, is in accord
with the spirit of the recently proposed universal relaxation bound
\cite{Hod1,Hod2}.

\bigskip
\noindent
{\bf ACKNOWLEDGMENTS}
\bigskip

This research is supported by the Meltzer Science Foundation. I
thank Liran Shimshi, Yael Oren and Clovis Hoppman for helpful
discussions.



\begin{thebibliography}{99}
\bibitem{Noteno} It is worth mentioning that all attempts to
decouple the gravitational and electromagnetic perturbations of the
Kerr-Newman spacetime have failed so far, see e.g. \cite{Chan}.
\bibitem{Notealm} In the small $a\omega$ limit we shall take
$A=l(l+1)-s(s+1)$ in Eq. (\ref{Eq3}), where $l$ is nearly an integer
with a small correction of order $O(a\omega)$.
\bibitem{Notevalid} Taking cognizance of Eq. (\ref{Eq10}), one finds that the total
reflection modes (TRM), which are characterized by the condition
${\cal T}(\omega)=0$, are given by the requirement
$1/\Gamma(-l-2i\varpi)=0$, that is $-l-2i\varpi=-n$. We recall that
$l$, as {\it defined} from the relation $A=l(l+1)-s(s+1)$ [where the
separation constants $\{A\}$ themselves are obtained from Eq.
(\ref{Eq2})], is nearly an integer with a small correction of order
$O(a\omega)$. This implies that each quasinormal frequency is
separated from a nearby total reflection frequency by a small term
of order $O(T_{BH}a\omega)$. Thus, in order to determine the QNMs in
numerical calculations (and distinguish them from the TRMs), one
would have to use numerical schemes of very high precision.
\end{thebibliography}
\end{document}